\documentclass[mathleft
]{an}
\usepackage{graphicx}
\usepackage{times}
\usepackage{natbib}
\newcommand{\aap}{    {\it Astron. Astrophys.}}

\def\kms{$\mathrm{km\, s^{-1}}$}

\newcommand{\COBOLD}{{\sf CO$^5$BOLD}}
\newcommand{\cobold}{\COBOLD}

\newcommand{\xx}{\ensuremath{\mathrm{1D}_{\mathrm{LHD}}}}
\newcommand{\mD}{\ensuremath{\left\langle\mathrm{3D}\right\rangle}}
\newcommand{\loggf}{\ensuremath{\log\,gf}}
\DeclareRobustCommand{\ion}[2]{%
\relax\ifmmode
\ifx\testbx\f@series
{\mathbf{#1\,\mathsc{#2}}}\else
{\mathrm{#1\,\mathsc{#2}}}\fi
\else\textup{#1\,{\mdseries\textsc{#2}}}%
\fi}

\begin{document}

\Pagespan{789}{}
\Yearpublication{2006}%
\Yearsubmission{2005}%
\Month{11}%
\Volume{999}%
\Issue{88}

\title{Sulphur abundances in halo stars from Multiplet\,3 at 1045\,nm\thanks{Using data from CRIRES at the ESO-VLT,
Programme 079.D-0434.}}

\author{E. Caffau\inst{1}\fnmsep\thanks{Corresponding author:
  \email{Elisabetta.Caffau@obspm.fr}\newline}
\and  L. Sbordone\inst{2}
\and  H.-G. Ludwig\inst{3}
\and  P. Bonifacio\inst{1,4}
\and  M. Spite\inst{1}
}
\titlerunning{Sulphur abundances in halo stars}
\authorrunning{E. Caffau el at.}
\institute{
GEPI, Observatoire de Paris, CNRS, Universit\'e Paris Diderot, Place
Jules Janssen, 92190
Meudon, France
\and 
Max-Planck Institut f\"ur Astrophysik, Karl-Schwarzschild-Str. 1, 85741, Garching, Germany.
\and 
Zentrum f\"ur Astronomie der Universit\"at Heidelberg, Landessternwarte, K\"onigstuhl 12, 69117 Heidelberg, Germany
\and
Istituto Nazionale di Astrofisica,
Osservatorio Astronomico di Trieste,  Via Tiepolo 11,
I-34143 Trieste, Italy}

\received{26 Feb 2010}
\accepted{10 Mar 2010}
\publonline{later}


\abstract{%
  Sulphur is a volatile $\alpha$-element which is not locked into dust grains
  in the interstellar medium (ISM).  Hence, its abundance does not need to be
  corrected for dust depletion when comparing the ISM to the stellar
  atmospheres.  The abundance of sulphur in the photosphere of metal-poor
  stars is a matter of debate: according to some authors, [S/Fe] versus [Fe/H]
  forms a plateau at low metallicity, while, according to other studies, there
  is a large scatter or perhaps a bimodal distribution.  In metal-poor stars
  sulphur is detectable by its lines of Mult.\,1 at 920\,nm, but this range is
  heavily contaminated by telluric absorptions, and one line of the multiplet
  is blended by the hydrogen Paschen $\zeta$ line.  We study the possibility
  of using Mult.\,3 (at 1045\,nm) for deriving the sulphur abundance because
  this range, now observable at the VLT with the infra-red spectrograph
  CRIRES, is little contaminated by telluric absorption and not affected by
  blends at least in metal-poor stars.  We compare the abundances derived from
  Multiplets 1 and 3, taking into account NLTE corrections and 3D effects.
  Here we present the results for a sample of four stars, although the
  scatter is less pronounced than in previous analysis, we cannot find a
  plateau in [S/Fe], and confirm the scatter of the sulphur abundance at low
  metallicity.}
\maketitle

\section{Introduction}

\sloppy
The light elements between O and Ti of even atomic number are referred to as
$\alpha$-elements because they are mainly produced by successively adding an
$\alpha$-particle from nucleus to nucleus.  Among them is sulphur, a volatile
element which is not locked into dust in the interstellar medium (ISM), so
that the sulphur abundance derived for the ISM can be directly compared to the
sulphur abundance derived in stars.

The $\alpha$-elements are crucial probes of the chemical
evolution of a stellar population: they are almost exclusively released by Type II Supernovae,
while the iron peak elements are produced by Type II SNe, 
but also, in large amounts, by Type Ia Supernovae.
Progenitors of Type II and Type Ia supernovae have very different lifetimes,
making the abundance ratio of $\alpha$-elements to iron-peak elements
a powerful diagnostic of the chemical evolution and star
formation history of a galaxy. 
In the Milky Way, stars of lower metallicity
are characterised by higher $\alpha$ to iron abundance ratios
than found in the Sun and stars of solar metallicity. This is
usually interpreted in terms of the lower contribution of Type Ia SNe.
Systems which are characterised by low or bursting star formation, 
like dwarf Spheroidal galaxies, give time to Type Ia SNe to explode
before the enrichment due to Type II SNe has greatly increased.
Consequently such systems display rather low $\alpha$ to iron ratios
even at low metallicities.
Thus the abundance of $\alpha$-elements is an important 
property of any stellar population.
For the study of chemical evolution in external galaxies,
the more readily available objects 
are Blue Compact galaxies (BCGs) through analysis
of the emission line spectra, and Damped Ly$-\alpha$
systems (DLAs) through the analysis of resonance absorption
lines. In both groups of objects, sulphur is relatively 
easy to measure in the form of ISM emission (for the BCGs) or
absorption (for the DLAs) lines.

The investigation of sulphur abundances in the stellar photospheres
started with the pioneering work of \cite{GeorgeW}. They
determined the sulphur abundance in six stars out of a sample of nine stars in six
Galactic Clusters. Later on, 
\cite{clegg81} determined the sulphur content in 20 F- G-type stars with [Fe/H]$\ge -1$,
while \cite{fran87,fran88} studied 13 and 12 metal-poor stars, respectively.

Not many sulphur lines are available in the observed stellar spectra.
There is a forbidden line from the ground level of \ion{S}{i} at 1082.1\,nm, 
which is weak, blended, but for which lower and upper level populations are
very close to local thermal equilibrium (LTE). 
This absorption line is
measurable in the spectrum of solar-like stars \citep{ryde06}.
However, it becomes undetectable below [Fe/H]$<-0.5$.
The permitted Mult.\,8 at 675\,nm is weak, but not blended, neither contaminated
by telluric absorption. For these transitions the assumption that high and low levels are
in LTE is a good approximation, 
and the same holds for the lines of Mult.\,6 at 869\,nm \citep{takada02}. 
Both Multiplets 6 and 8 are weak,
as a consequence detectable only in stars of solar, or moderately sub-solar metallicity 
(down to [Fe/H]$\sim -1.5$).
Below such metallicity, usually Mult.\,1 at 920\,nm is used, which is, on the other hand, 
contaminated by telluric absorption. Moreover one of the lines
of the triplet (922.8\,nm) is located in the blue wing of the hydrogen
Paschen $\zeta$ line,
and the presence of the hydrogen line makes the abundance determination more difficult.

The infra-red lines of Mult.\,3 at 1045\,nm are well suited to measure the
sulphur abundance. These lines are not as strong as the
components of Mult.\,1, but easily detectable in very metal-poor stars.
No blend has to be taken into account in the case of metal-poor stars
because the only extra line present in the range, the \ion{Fe}{i} line at 1045.5\,nm 
blending the strongest \ion{S}{i}
line of the triplet, vanishes at sub-solar metallicity.
Telluric absorptions in this range are less abundant and less strong than in the case of Mult.\,1.
Observing the 1045\,nm sulphur lines provides thus a possibility
to obtain a reliable sulphur abundance in very metal-poor stars.

\cite{zolfo} analysed the sulphur abundance in a sample of Galactic stars. 
They investigated, when available, the lines of Multiplets 6, 8, and 1.
Because of the blend of the 922.8\,nm \ion{S}{i} line of Mult.\,1  with the
hydrogen Paschen $\zeta$, they performed line profile fitting to derive the sulphur content
for all the lines available.
The line profile fitting procedure permitted to reproduce the line profile
of the hydrogen. In their study they 
have suggested that in the range $-2.5<$[Fe/H]$<-2.0$ the [S/Fe] ratio
shows either a large scatter or a bimodal behaviour.
Most of the stars lie on a ``plateau'' at about [S/Fe]=+0.4, while
a non negligible number of stars shows a ``high'' value
of [S/Fe], around +0.8.
This behaviour has no proposed interpretation, but
it may be supposed to be due to systematic errors affecting only
the analysis of Mult.\,1.
In fact, in the sample of \cite{zolfo}, the determination
of [S/Fe] in this range of metallicity is based mainly on the
non contaminated lines of Mult.\,1.
On the other hand, 
\cite{nissen04} and \cite{nissen07}
derived the sulphur abundance from equivalent width (EW) measurements of the
lines of multiplets 8 and 1, and they
find a plateau at low metallicity
in the [S/Fe] versus [Fe/H] plot, with no sign of bimodal distribution or scatter.
It is therefore of great interest to verify this puzzling
finding by the use of an independent and, probably better,
diagnostic of the sulphur abundances, as can be afforded
by the 1045\,nm lines.

\section{Model atmospheres and atomic data}

This analysis was performed by means of 1D hydrostatic model atmospheres. 
We then applied NLTE corrections taken from \cite{takeda05}
and 3D corrections derived by means of \cobold\ 3D hydrodynamical models 
\citep{freytag02,wedemeyer03}.
The 1D model atmospheres used to derive the sulphur abundance 
were obtained from the Linux version of ATLAS, as described
in \cite{sbordone04}.
To derive the 1D-LTE abundance of sulphur 
we measured the equivalent widths (EW) with the IRAF task {\tt splot}
and derived abundances through WIDTH  \citep[for details see][]{kurucz93,kurucz05,castelli05}.

The 3D-\cobold\ models used to compute the 3D-corrections
(listed in Table~\ref{mod3d})
are taken from the CIFIST grid \citep[][]{grid}.
For each 3D model, we used as reference model
the 1D model obtained as the horizontal average of each 3D snapshot
over surfaces of equal (Rosseland) optical depth, and
the 1D hydrostatic model computed with the LHD code
with the same atmospheric parameter as the 3D model.
Both these 1D models share the micro-physics and radiative transfer scheme with \cobold.
A description of these models can be found in \cite{zolfito}.
We computed 3D corrections as described in \cite{zolfito} and in 
\citet{caffaunew}.

The 3D corrections are given in Table~\ref{sabbo} in the case of a
micro-turbulence of 1.5\kms\ used in the 1D reference model.
${\Delta^{(1)}(\xi_{\rm mic})=A{\rm (S)}_{\rm 3D}}-A{\rm (S)}_{\rm \mD}$ in
column 7 takes into account the effects of granulation, meaning the influence
that the horizontal fluctuations around the mean stratification have on the
abundance determination.
${\Delta^{(2)}(\xi_{\rm mic},\alpha_{\rm MLT})=A{\rm (S)}_{\rm 3D}}-A{\rm (S)}_{\rm \xx}$
in column 8 accounts for the total 3D effect, meaning both the effect of horizontal fluctuations
and the different mean temperature structure due to the different treatment of 
convection in 3D hydrodynamical models and
in 1D mixing-length models.

\begin{table*}
\caption{Hydrodynamical model atmospheres}
\label{mod3d}
\begin{tabular}{cccccc}
\hline
Model & Teff & log g & [Fe/H] & $\Delta$t & Box-size\\ 
      &  K   &       &        & h         & km$^3$ \\
\hline
d3t50g35mm10n01 & 4930 & 3.5 & --1.0 & 117 & $59732.2 \times 59732.2 \times 30199.5$ \\
d3t50g35mm20n01 & 4980 & 3.5 & --2.0 & 151 & $59732.3 \times 59732.3 \times 30199.5$ \\
d3t50g45mm10n03 & 5060 & 4.5 & --1.0 &  18 & $5084.46 \times 5084.46 \times 2477.64$ \\
d3t59g35mm20n01 & 5860 & 3.5 & --2.0 &  31 & $59308.6 \times 59308.6 \times 38658.3$ \\
d3t59g35mm30n01 & 5870 & 3.5 & --3.0 &  31 & $59308.6 \times 59308.6 \times 38658.3$ \\
d3t59g45mm20n01 & 5920 & 4.5 & --2.0 &   7 & $~~6020.0 \times ~~~6020.0 \times 3841.66$ \\
 \\
\hline
\end{tabular}
\end{table*}

The atomic data of the \ion{S}{i} lines of Mult.\,3, the same as in
\cite{smult3}, are summarised in Table\,\ref{adata}.

\begin{table}
\caption{\ion{S}{i} lines of Mult.\,3}
\label{adata}
\begin{tabular}{ccrc}
\hline
\\
$\lambda$ & Transition               & \loggf  & $\chi _{\rm lo}$  \\
(nm) air  &                          &         &              (eV)\\
\noalign{\smallskip}
\hline
\noalign{\smallskip}
1045.5449  & $^3\mathrm{S}_1^\mathrm{o}-{^3}\mathrm{P}_2$ &   0.26  & 6.86 \\
1045.6757  & $^3\mathrm{S}_1^\mathrm{o}-{^3}\mathrm{P}_0$ & --0.43  & 6.86 \\
1045.9406  & $^3\mathrm{S}_1^\mathrm{o}-{^3}\mathrm{P}_1$ &   0.04  & 6.86 \\
 \\
\hline
\end{tabular}
\end{table}

\section{Analysis of CRIRES data}

We observed four metal-poor dwarf stars with CRIRES \citep{kaufl04}.
The observations, programme
079.D-0434 (E. Caffau), have been carried out in service mode,
for details see Table~\ref{odata}.
The spectra were observed using detector integration times
of 30\,s or 10\,s (for the brighter stars), the total integration
time for each star is given in Table~\ref{odata}.
The observation template employed, required nodding along
the slit, with an amplitude of 10$''$.
The A0 star HIP 28910 was also observed, to check the
position of telluric lines.

The analysis was conducted using the extracted, 
combined and calibrated spectra provided by ESO. We compared this
data with that obtained for one star by running a set of reduction routines based on
the ESO CRIRES pipeline 1.6.0, but did not find any significant
difference with respect to the reduced data provided by ESO.

\begin{table}
\caption{\ion{S}{i} lines of Mult.\,3}
\label{odata}
\begin{tabular}{lcrr}
\hline
\\
 Star        & J    & date       & total integration time \\
             & mag  &            &  s           \\
\noalign{\smallskip}
\hline
\noalign{\smallskip}
\hbox{BD\,$-05^\circ 3640$}  & 7.62 & 24-07-2007 &  2$\times$ 2160\\
HD 140283    & 6.01 & 24-07-2007 &  300 \\
HD~165195    & 4.89 & 24-07-2007 &  400 \\
HD~181743    & 8.62 & 19-06-2007 &  2160 \\
             &      & 23-07-2007 &  2160 \\
HD~211998    & 4.15 & 22-07-2007 &  160  \\
\\
\hline
\end{tabular}
\end{table}

Our sample consisted of four Galactic stars from the sample of \cite{zolfo}.
Three of them, \hbox{BD\,$-05^\circ 3640$},
HD~181743, and HD~211998,
have been analysed by \cite{zolfo}, and
the sulphur abundance derived is the one in Table\,\ref{sabbo}.
HD~181743 has also been analysed by \cite{nissen04}. As stellar parameter
T$_{\rm eff}$/$\log{\rm g}$[Fe/H] they determined 5863\,K/4.32/--1.93, and
their sulphur abundance was A(S)=5.61, in agreement, within errors
with our result..
HD~211998 has also been analysed by \cite{israelian01} and \cite{fran87}.
\cite{israelian01} derived A(S)=5.80 
with the stellar parameters T$_{\rm eff}$/$\log{\rm g}$/[Fe/H] of
5271\,K/3.36/--1.25. The difference of more than 0.3\,dex with respect to 
our result is anyway compatible within errors, in fact for this star their error is about
0.3\,dex. We stress also a difference of 0.3\,dex in metallicity.
Our analysis is in close agreement with the one of \cite{fran87}, who finds A(S)=6.20.
For HD~140283 the values in Table\,\ref{sabbo} are from \cite{nissen04}.

Our goal was to understand if Mult.\,3 can be used in order to
derive the sulphur abundance in metal-poor stars. It could be that the large scatter
in [S/Fe] at low metallicity described in \cite{zolfo} is due to systematic effects
arising from the use of one specific \ion{S}{i} multiplet for abundance determination.
A sample of 4 stars is insufficient to firmly establish whether this is the case or not, but
it can provide useful insights to direct further research.

The stellar parameters for the stars of our sample are taken from \cite{zolfo} and are 
summarised in columns 2-4 of Table~\ref{sabbo}.
In column~5 the 1D-LTE abundance with the
uncertainty related to the EW measurement is given.
While in \cite{zolfo} no NLTE correction was applied, here NLTE corrections
are interpolated in the table of \cite{takeda05} and are listed in column~6 of Table~\ref{sabbo}.
The 3D corrections for each star are given in column~7 and 8.
In column~9 [S/Fe] is given;
both NLTE and ${\Delta^{(2)}}$ 3D corrections are applied.
We adopt A(S)=7.16 for the solar sulphur abundance, 
as described in \citet{caffaunew}.
In column~10 the sulphur abundance from \cite{zolfo} is reported,
and in column~11 there is the relative NLTE correction from \cite{takeda05}.
The comparison in [S/Fe] is not straightforward. To derive the [S/Fe] given in column~9
of Table~\ref{sabbo}, both NLTE and 3D-$\Delta^{(2)}$ corrections are applied.
As one can see in the Table, these two corrections roughly cancel each other,
so that the ${\rm 1D_{\rm LTE}}$ abundance we derive is close to the abundance
obtained after the application of NLTE and 3D corrections.
Another difference with respect to \cite{zolfo} is the solar reference abundance, 
${\rm A(S)_\odot}=7.16$ in this analysis, while \citet{zolfo} adopted 
${\rm A(S)_\odot}=7.21$ from \cite{anders89}.
For this reasons we suggest to compare the A(S) 1D-LTE obtained in this work
and in \cite{zolfo}, in column~4 and 10 of Table~\ref{sabbo} respectively.

\begin{table*}
\caption{Stellar parameters and sulphur abundances. 
To derive [S/Fe] in column~9, NLTE and 3D corrections have been applied. For the solar reference we adopt ${\rm A(S)}_\odot=7.16$.}
\label{sabbo}
\begin{tabular}{lcccccccccc}\hline
Star & Teff & $\log{\rm g}$ & [Fe/H] & A(S)-1D\,LTE &{$\Delta_{\rm LTE}$} & $\Delta^{(1)}$ & $\Delta^{(2)}$ & [S/Fe] & A(S)$^{\rm C05}$ & $\Delta_{\rm LTE}$\\ 
     & K    &        &        & & Mult.\,3 & & & & \\
 (1) & (2) & (3) & (4) & (5) & (6) & (7) & (8) & (9) & (10) & (11)\\
\hline
\hbox{BD\,$-05^\circ 3640$} 
             & 5020 & 4.61 & --1.19 & $6.61\pm 0.10$ & --0.02 & 0.03 & 0.05 & $0.67\pm 0.14$ & $6.77~\pm 0.10$ & --0.03 \\
HD~140283    & 5690 & 3.69 & --2.42 & $5.12\pm 0.05$ & --0.15 & 0.00 & 0.08 & $0.31\pm 0.11$ & $5.10~\pm 0.10$ & --0.16 \\ 
HD~181743    & 5970 & 4.40 & --1.81 & $5.73\pm 0.10$ & --0.07 & 0.02 & 0.13 & $0.44\pm 0.14$ & $5.83^{\rm u}\pm 0.10$ & --0.04 \\
HD~211998    & 5210 & 3.36 & --1.56 & $6.15\pm 0.10$ & --0.12 & 0.06 & 0.12 & $0.55\pm 0.14$ & $6.33~\pm 0.10$ & --0.19 \\ 
 \\
\hline
\end{tabular}
\\
\\
${\Delta^{(1)}(\xi_{\rm mic})=A{\rm (S)}_{\rm 3D}}-A{\rm (S)}_{\rm \mD}$\\
${\Delta^{(2)}(\xi_{\rm mic},\alpha_{\rm MLT})=A{\rm (S)}_{\rm 3D}}-A{\rm (S)}_{\rm \xx}$\\
u: upper limit from this work\\
C05: \cite{zolfo}
\end{table*}

\begin{figure}
\includegraphics[width=80mm,clip=true]{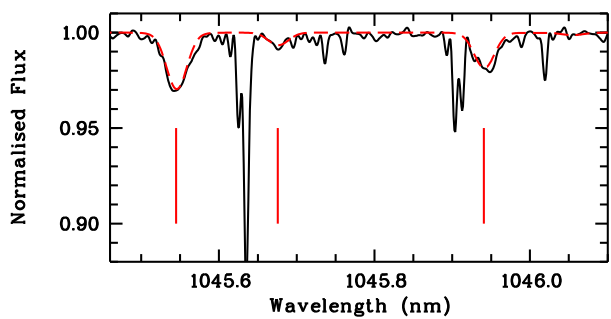}
\caption{The spectral region of the \ion{S}{i} lines of Mult.\,3 in the case of
the metal-poor star HD\,140283 (solid black) over-imposed to an
ATLAS+SYNTHE synthetic spectrum (5690K/3.69/–2.5, A(S)=5.11, dashed red).}
\label{hd140283}
\end{figure}

For star \hbox{BD\,$-05^\circ 3640$}, the ``high'' sulphur abundance 
derived in \cite{zolfo} is here confirmed. The difference
of 0.16\,dex between the two analysis is large but within the errors. 
The same is true for HD\,211998: for this star the NLTE correction
for Mult.\,1 is 0.07\,dex larger than for Mult.\,3.
For HD\,140283 the two analyses are in good agreement.
In Fig.\,\ref{hd140283} we show the observed spectrum (solid-black line)
of HD\,140283 compared to an ATLAS+SYNTHE synthetic spectrum (dashed-red line).
For star HD\,181743 we find a difference between this analysis
and the results of \cite{zolfo} larger than expected from the error bars.
The previous analysis relied on Mult.\,6, the observed spectrum was 
observed with UVES@VLT, with resolution of 43\,000, dichroic \#2 
and  cross disperser \#4. In this spectrum a line was clearly visible
at the wavelength of the Mult.\,6, with a FWHM compatible with the iron line close by.
The spectrum showed no fringing as suggested by \cite{nissen07}.
A possible explanation for the presence of this line is that it is a contribution
of the sky, not perfectly subtracted.
In fact at the time of that observation the Moon was bright (68\% illumination)
and close to the star ($29^\circ$ distance).
Due to the large discrepancy we found with the analysis of the lines of Mult.\,3,
we reduced the archival UVES spectra of this star taken during ESO programmes 67.B-0474(A) 
(P. I. Holmberg) as well as
67.D-0106(A) (P.I. Nissen). In the first programme a single 600',s, R$\sim$60000 
spectrum was produced with CD\,\#3, while in the second case a 
set of 400\,s, R$\sim$60000 spectra were produced with image slicer \#1 and CD\,\#2.
With these data we cannot detect the Mult.\,6 and are only able to give an upper limit 
for the sulphur abundance, which is reported in Table~\ref{sabbo}.
We confirm the result of \cite{nissen07} about the Mult.\,6, and find a sulphur abundance
from Mult.\,3 very close to their value. In fact, according to their LTE analysis, [S/Fe]=0.29
with ${\rm A(S)}_\odot=7.20$, while we find [S/Fe]$=0.38\pm 0.010$ with ${\rm A(S)}_\odot=7.16$.

\section{Discussion}

\begin{figure}
\includegraphics[width=80mm,clip=true]{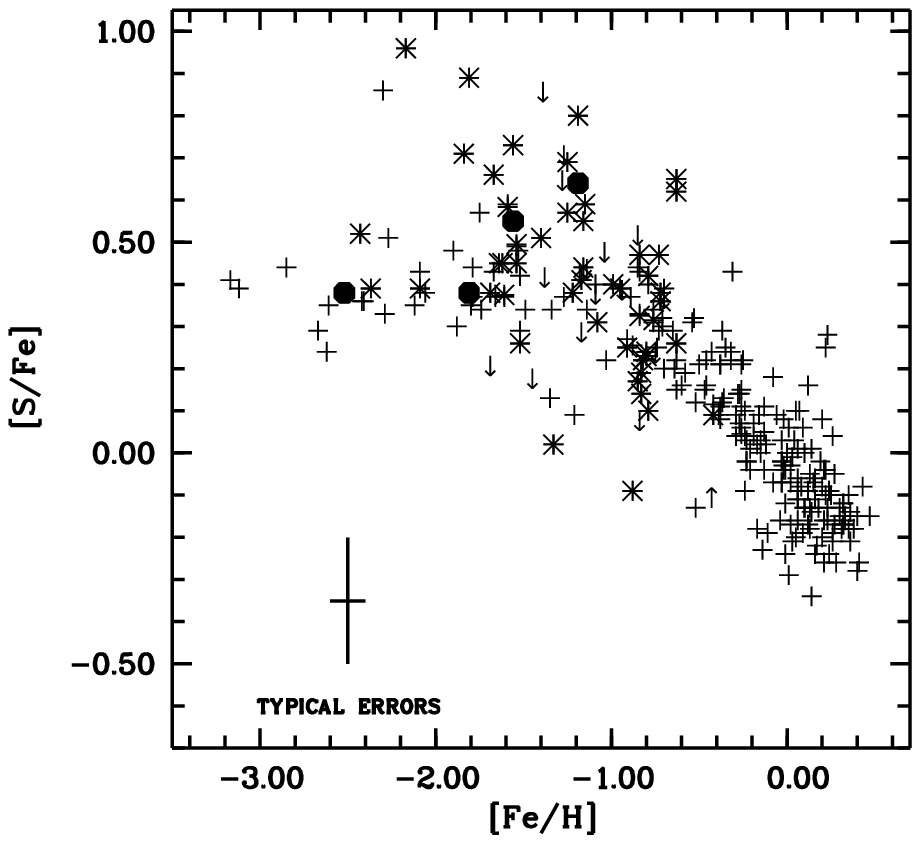}
\caption{The 1D-LTE results of \cite{zolfo}, from their analysis (asterisks)
and taken from the literature (crosses),
meaning from \cite{israelian01,chen02,takada02,ecuvillon04,ryde04,nissen04}, 
are compared
to our value (filled circles) from Mult.\,3.}
\label{zolfo05}
\end{figure}

\cite{zolfo} derived the sulphur abundance in a sample of metal-poor stars
and included in their analysis the results available in the literature for a large number
of Galactic stars.
Their complete sample spans a wide range in metallicities, 
from solar to 1/1000 of the solar metallicity.
As visible in their figure~10, 
the S abundance, as a function of the stellar iron content, has a negative
trend for stars of solar metallicity or for slightly metal-poor stars
($-1.0<$[Fe/H]$<+0.5$).
For low metallicity ([Fe/H]$<-1.5$) a large scatter or a bimodal behaviour is visible.
This trend, which at the moment has no physical explanation,
could have been attributed either to the fact that 
the sulphur abundance in metal-poor stars was mainly derived from the \ion{S}{i} lines of 
Mult.\,1 where telluric absorption can be a problem, or to
the neglect of NLTE effects.
NLTE effects are not very large, about --0.1\,dex, and in part they are
compensated by 3D effects.
The scatter we find in this work from Mult.\,3 from our 4 stars is larger than 0.3\,dex,
compatible only within 2$\sigma$ with a plateau.
In Fig.\,\ref{zolfo05} our 1D-LTE results are compared to the results of \cite{zolfo},
which have been renormalised to ${\rm A(S)_\odot}=7.16$.

\cite{smult3} investigated the sulphur abundance derived from
the \ion{S}{i} lines of Mult.\,3 in the Sun and a small sample of solar metallicity stars.
They compared this abundance with the one obtained from 
the weaker lines of Multiplets 6 and 8. Their analysis
was carried out before CRIRES was available and the needed spectral range was obtained by
forcing the extraction of the last order in the UVES 860\,nm standard setting of a few bright,
solar-metallicity stars taken from the Paranal Observatory Project dataset \citep{bagnulo03}.
Except for the Sun, the abundances obtained from the
different lines of \ion{S}{i} are in agreement.

\cite{takada02} derived sulphur abundances in six giants and 61 dwarfs, 
in the range $-3\le[{\rm Fe/H}]\le+0.5$, using Mult.\,6, and found a linear increase of 
[S/Fe] versus [Fe/H] with decreasing metallicity.

\cite{takada05} analysed Multiplets 6 and 1 in 21 metal-poor stars.
They find a good agreement of A(S) from the two multiplets for [Fe/H]$> -2$,
while for the most metal-poor stars the abundances from Mult.\,6 are larger
than from Mult.\,1.
In their analysis [S/Fe] as a function of the metallicity forms a plateau.

\begin{figure}
\includegraphics[width=80mm,clip=true]{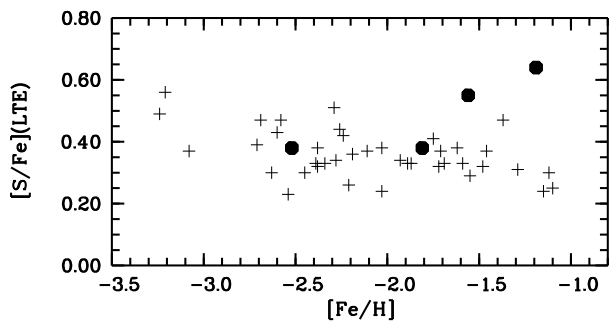}
\caption{The 1D-LTE results of \cite{nissen07} (crosses) are compared
to our value (filled circles) from Mult.\,3.}
\label{sfe_fe}
\end{figure}

\cite{nissen07} observed Mult.\,3 of G~29-23, using CRIRES science verification data,
and obtained a good agreement of the LTE sulphur abundance with the result 
obtained from the other lines.
In Fig.\,\ref{sfe_fe} the sample of \cite{nissen07}, scaled by 0.04\,dex for
the difference in the solar sulphur between the two analysis, is compared to our results.
The two most metal-rich stars we studied show a [S/Fe] larger than the values
in \cite{nissen07} for similar metallicity, while the other two stars fall on the average.
A scatter of about 0.3\,dex in [S/Fe] for metal-poor stars is comparable to the
scatter in [Mg/Fe] available in \cite{gratton03} for a similar range in metallicity
and in \citet{andrievsky10} for more metal-poor stars.

\section{Conclusions}

We did an analysis on the \ion{S}{i} lines of Mult.\,3 for four metal-poor
dwarfs stars of the sample of \cite{zolfo}.  The results of three out of the
four candidates are consistent with the previous analysis within errors, and
for these three stars the previous analysis relied on Mult.\,1.  We have to
notice that for three stars the sulphur abundance we obtain from the lines of
Mult.\,3 is lower with respect to the previous analysis based on Mult.\,1.
For HD\,181743, whose previous analysis was based on Mult.\,6,
the difference of 0.5\,dex is above the expected error, and
can now be explained with a contamination from the sky in the spectrum
analysed by \cite{zolfo}.

From the results of this small sample we cannot confirm the existence of a
plateau in the [S/H],[Fe/H] plane. The scatter is larger with respect to
the estimated uncertainties, but a similar scatter is present in [Mg/Fe] versus [Fe/H]
in the analysis of \cite{gratton03}.
The systematic error related to the oscillator strength of the lines
of Mult.\,1 can be
neglected because it acts in the same way for the complete sample of stars.
One has to take into account the uncertainties related to the temperature,
but these error cannot remove the great scatter in [S/Fe]. 
In fact for the cooler star a change of 100\,K
in the effective temperature is translated in a change of 0.10\,dex in the
sulphur abundance, while for the hotter star the same difference in temperature
would produce only 0.04\,dex change in A(S).

An analysis on Mult.\,3 in an extended sample of stars would be useful to
investigate the behaviour of [S/Fe] at low metallicity.


\acknowledgements
  We acknowledge use of the supercomputing centre CINECA,
  which has granted us time to compute part of the hydrodynamical
  models used in this investigation, through the INAF-CINECA
  agreement 2006,2007,2008.



\end{document}